\documentclass[twocolumn, useAMS,usenatbib]{mn2e}%

\usepackage{amsmath}
\usepackage{graphicx}
\usepackage{lscape}
\usepackage{multirow}
\usepackage{multicol}
\usepackage{deluxetable}

\newcommand{\ssun}{\odot}
\usepackage{afterpage}

\newcommand{\lesssim}{\la}
\newcommand{\gtrsim}{\ga}

\newcommand{\fdegr}{.\!\!^\circ}

\newcommand{\lmc}{{\rm LMC X--1}}

\newcommand{\simpl}{{\sc simpl}}

\newcommand{\tbvarabs}{{\sc tbvarabs}}

\newcommand{\simplr}{{\sc simpl-R}}
\newcommand{\simplc}{{\sc simpl-C}}
\newcommand{\ireflect}{{\sc ireflect}}
\newcommand{\reflionx}{{\sc reflionX}}
\newcommand{\reflionX}{{\sc reflionX}}
\newcommand{\reflion}{{\sc reflion}}

\newcommand{\relconvf}{{\sc relconvf}}
\newcommand{\refbhb}{{\sc refbhb}}

\newcommand{\refbhbm}{{\sc refbhb-M}}

\newcommand{\kerrbb}{{\sc kerrbb}}
\newcommand{\diskbb}{{\sc diskbb}}
\newcommand{\kerrbbtwo}{{\sc kerrbb2}}
\newcommand{\bhspec}{{\sc bhspec}}

\newcommand{\Mdot}{\dot{M}}

\newcommand{\msun}{\rm M_{\ssun}}

\newcommand{\rchinu}{\chi^{2}/\nu}
\newcommand{\chisq}{\chi^{2}}
\newcommand{\fsc}{f_{\rm SC}}
\newcommand{\nh}{N_{\rm H}}

\newcommand{\cm}{\rm cm}

\newcommand{\ka}{{$K\alpha$}}
\newcommand{\spin}{a_{*}}

\newcommand{\lledd}{L_{\rm X}/L_{\rm Edd}}

\newcommand{\rxte}{{\it RXTE}}

\newcommand{\suzaku}{{\it Suzaku}}

\title[A Broad Iron Line in LMC X--1]{A Broad Iron Line in LMC X--1}

\author[Steiner et al.]{James F.\ Steiner$^{1,2}$\thanks{E-mail: jsteiner@ast.cam.ac.uk},
  Rubens C.\ Reis$^{3}$,
  Andrew C. Fabian$^1$, Ronald
  A. Remillard$^4$,\newauthor Jeffrey E.\
  McClintock$^2$, Lijun Gou$^{2,5}$, Ryan
  Cooke$^1$, Laura W. Brenneman$^2$, \newauthor and Jeremy
  S.\ Sanders$^1$\\
$^{1}${Institute of Astronomy, Cambridge University, Madingley Road, Cambridge, CB3 0HA}\\
$^{2}${Harvard-Smithsonian Center for Astrophysics, 60 Garden Street, Cambridge, MA 02138, USA}\\ 
$^3${Dept. of Astronomy, University of Michigan, Ann Arbor, Michigan 48109 USA}\\
$^{4}${MIT Kavli Institute for Astrophysics and Space Research, MIT, 70 Vassar Street, Cambridge, MA 02139, USA}\\
$^5${National Astronomical Observatories, CAS, 20A Datun Road, Chaoyang District, 100012 Beijing, China}
}

\begin{document}

\maketitle

\begin{abstract}
  We present results from a deep \suzaku\ observation of the black
  hole in LMC X--1, supplemented by coincident monitoring with \rxte.
  We identify broad relativistic reflection features in a soft
  disc-dominated spectrum. A strong and variable power-law component
  of emission is present which we use to demonstrate that enhanced
  Comptonisation strengthens disc reflection.  We constrain the spin
  parameter of the black hole by modelling LMC X--1's broad reflection
  features. For our primary and most comprehensive spectral model, we
  obtain a high value for the spin: $\spin = 0.97^{+0.01}_{-0.13}$ (68
  per cent confidence).  However, by additionally considering two
  alternate models as a measure of our systematic uncertainty, we
  obtain a broader constraint: $\spin = 0.97^{+0.02}_{-0.25}$.  Both
  of these spin values are entirely consistent with a previous
  estimate of spin obtained using the continuum-fitting method.  At
  99\% confidence, the reflection features require $\spin > 0.2$.  In
  addition to modelling the relativistically broadened reflection, we
  also model a sharp and prominent reflection component that provides
  strong evidence for substantial reprocessing in the wind of the
  massive companion.  We infer that this wind sustains the ionisation
  cone surrounding the binary system; this hypothesis naturally
  produces appropriate and consistent mass, time, and length scales
  for the cone structure.

\end{abstract}


\section{Introduction}\label{section:Intro}

LMC X--1 was the first extragalactic black hole (BH) binary to be
discovered; Cygnus X--1 is the only other such persistent X-ray source
with an O-giant companion that is located locally, i.e., within the
Galaxy and Magellanic Clouds.  \lmc\ is quite unusual in that it
consistently maintains a stable luminosity of $\lledd \approx 16$ per
cent \citep{Gou_2009} despite showing strong fluctuations in the rms
amplitude of its power spectrum.

Recently, the spin\footnote{Defined as the dimensionless parameter
  $\spin \equiv cJ/GM^2$ with $|\spin| \leq 1$, where $M$ and $J$ are
  the BH's mass and angular momentum.} of \lmc\ was measured by
modelling the thermal accretion disc emission \citep{Gou_2009} via the
X-ray continuum-fitting technique (e.g., \citealt{Zhang_1997}).  Using
a primary sample of 18 \rxte\ PCA spectra, \citeauthor{Gou_2009}
estimated the spin to be $\spin=0.92^{+0.05}_{-0.07}$.

The principal alternative to continuum fitting is the reflection (or
Fe-line) method.  Here, the breadth of relativistically broadened
reflection features generated in the accretion disc, most notably the
prominent Fe \ka\ line complex, is used to determine spin (e.g.,
\citealt{Fabian_1989, Brenneman_Reynolds, miller07review}).  One
advantage of the Fe-line method over X-ray continuum fitting is that
it can be readily applied to measure the spins of stellar-mass and
supermassive BHs alike.  In the case of stellar-mass BHs, the Fe-line
method has the further virtue that it is independent of BH mass and
distance; however, its primary drawback is that it relies upon a much
fainter signal.  Recent studies (e.g., \citealt{Miller_2009,
  Steiner_j1550spin}) have made headway in achieving measurements via
both techniques.

Both methods make one fundamental assumption: that the accretion disc
is truncated at the innermost stable circular orbit (ISCO).  The
radius of the ISCO is uniquely defined by the BH's mass and spin,
growing with increasing mass, and shrinking with increasing spin.  The
ISCO-truncation assumption is presently at the forefront of
theoretical scrutiny, but is generally supported by
magnetohydrodynamic simulations (e.g., \citealt{Reynolds_Fabian_2008,
  Shafee_2008, Penna_2010, Kulkarni_2011, Noble_2011, Zhu_2012,
  Schnittman_2012}; but see \citealt{Noble_2009}, and references
therein).  At the same time, several empirical studies of BH binaries
in disc-dominated states provide strong support for the existence of a
stable inner-disc radius, including a comprehensive study analysing
decades of spectra of LMC X--3 \citep{Steiner_2010} and other studies
of BH spectral evolution during outburst (e.g., \citealt{Done_2007,
  Gilfanov_2010}).

Fortuitously, BH X-ray binary sources located in the LMC are in
\suzaku's `Goldilocks zone' for Fe-line observations.  By virtue of
being located at $\sim$50~kpc distance, even luminous BHs are faint
enough that pileup effects are
modest\footnote{\citet{Miller_2010_pileup} discuss pileup and its
  impact on line fitting.}, while at the same time these sources are
bright enough to provide the signal-to-noise required for reflection
spin measurements in a reasonable observation time of $\sim$100~ks.
By comparison, pileup in standard \suzaku\ observing modes is severe
for outbursting Galactic sources, while Fe-line spin measurements for
practical observing times would be photon starved for stellar-mass BHs
beyond the Magellanic Clouds.

LMC X--1 is propitious for study because, despite being in an almost
persistently disc-dominated state, its spectrum contains a relatively
strong power-law component (that is required to produce reflection).
Tentative evidence for broad Fe \ka\ emission (with equivalent width
$\approx200$~eV) was reported by \citet{Nowak_2001}.  We unambiguously
confirm the presence of a broad Fe line in LMC X--1, and we use it
along with other reflection features to constrain the BH's spin.  We
thereby demonstrate that Fe fluorescence emission is present in the
soft states of \lmc\ and we furthermore show that the strength of this
feature is directly related to the strength of the Compton power-law
component.

Our \suzaku\ and \rxte\ observations of LMC X--1 are described in
Section~\ref{section:data} and the spectral models we employ are
described in Section~\ref{section:model}.  Our analysis is presented
in Section~\ref{section:results} followed by a discussion in
Section~\ref{section:discussion}, and we offer our conclusions in
Section~\ref{section:conc}.

\section{Observations}\label{section:data}

We observed \lmc\ for 130~ks using \suzaku\ from 2009 July 21 through
2009 July 24 with the XIS detectors operating in quarter-window mode.
The data were reduced following the XIS and PIN pipeline procedures.
For the XIS, we found it necessary to increase the {\it sisclean}
threshold to prevent contamination of the image core\footnote{See
  http://www-utheal.phys.s.u-tokyo.ac.jp/}.  During the observation,
the source count rate was stable, varying from its average intensity
by less than 5 per cent.  Background spectra were obtained from
observations of the Lockman Hole.  In this and other reduction steps,
we have been guided by the procedures described in
\citet{Kubota_2010}.

\suzaku's attitude calibration was improved using the {\it aeattcor}
routine\footnote{http://space.mit.edu/CXC/software/suzaku/aeatt.html}
\citep{Nowak_2011}.  The innermost region of the point-spread function
(PSF) suffered from moderate ($\sim10$ per cent) photon pileup.  To
ameliorate this problem, we used the utility {\it
  pile\_estimate}\footnote{http://space.mit.edu/CXC/software/suzaku/pest.html}
as a guide, and excised the innermost 30$\arcsec$.  Doing so, we
retained $\approx70$ per cent of the flux, while keeping the net
pileup below 3 per cent.

Three XIS units were used to collect the data: the two
front-illuminated detectors, XIS-0 and XIS-3, and the back-illuminated
detector, XIS-1.  Spectra were binned to approximately half \suzaku's
energy resolution and analysed over 0.8--10~keV (XIS-0,3); 0.8--8 keV
(XIS-1).  Because of calibration defects, we omitted channels between
1.5--2.5 keV and added a Gaussian line near 3.2 keV to model a
calibration glitch \citep{Kubota_2010}.  We also included a narrow
Gaussian line near 0.82 keV which was allowed to have separate
normalisations between the front and back-illuminated units.
Following Kubota et al., a 1 per cent systematic uncertainty was added
to all XIS energy channels.  PIN data were analysed from 18--50~keV,
while using the `tuned' instrumental background \citep{Fukazawa_2009}.
The cosmic X-ray background (CXB) contribution to the PIN
spectrum\footnote{The CXB contributes negligibly to the XIS.} is
included as a spectral model
component\footnote{http://heasarc.nasa.gov/docs/suzaku/analysis/pin\_cxb.html}
with 10 per cent freedom in its normalisation.  We fitted for the
normalisation of the instrumental background spectrum, which has a
nominal uncertainty of $3$ per
cent\footnote{http://heasarc.gsfc.nasa.gov/docs/suzaku/analysis/abc/\label{suznote}}.

During the \suzaku\ observation, we obtained eleven \rxte\
pointings\footnote{To avoid contamination from the nearby PSR
  B0540--69 \citep{Haardt_2001}, \rxte\ was offset by $0\fdegr25$ in
  the opposite direction, and we have generated an off-axis response
  calibration to correct for this, while including a 1 per cent
  systematic uncertainty in each spectrum.} (ranging from ~1--11 ks
apiece), which are interspersed throughout our \suzaku\ observing
window and bracket it.  We exclusively used the best-calibrated,
`standard 2' spectra from PCU-2.  These data improved the constraints
on the continuum features, including the high-energy power-law
component and the Compton hump.  The \rxte\ spectra were background
subtracted, corrected for detector dead time, and analysed from
2.55--45~keV.

As in previous work (see \citealt{Steiner_2010, Steiner_j1550spin}),
we standardised the shape and normalisation of each detector's
calibration so that the power-law parameters of Crab spectra match the
\citet{Toor_Seward} values; we introduce a floating
cross-normalisation between detectors to account for any residual
difference ($\lesssim 5$ per cent), except for the PIN which is
assigned a fixed normalisation of 1.16 relative to XIS-0.  Motivated
by \citet{Tsujimoto_2011} and \citet{Ishida_2011}, which show the
back-illuminated XIS detector yields a significantly different
spectral index than the two front-illuminated detectors, we include an
extra parameter for the difference between the XIS-1 spectral index
and those of XIS-0/3.  This is incorporated into the model as a fit
parameter and found to be quite modest: $\Delta\Gamma \approx 0.015
\pm 0.005$ (Table~\ref{tab:results}).  The additional freedom in the
model significantly improves our fits ($\Delta\chisq \approx 25$), and
insignificantly alters the value of spin (the change is $\lesssim 10$
per~cent of the statistical uncertainty).

\section{Models}\label{section:model}

We introduce two similar models, which we refer to below as Model~1
and Model~2, that differ in our treatment of the reflection component.
Both models assume that this component is generated in the same
manner: by disc photons that are Compton scattered in the corona and
then reprocessed in the accretion disc.  In the first model, we employ
the {\sc reflion} \citep{reflionx} family of reflection spectral
models, and in the second, use the {\sc pexrav} \citep{IREFLECT}
family.  In both cases, we model the thermal disc emission using
\kerrbbtwo\ \citep{KERRBB,BHSPEC, McClintock_2006} and the Compton
component via \simplr\ \citep{Steiner_simpl, Steiner_j1550spin}.

For the \reflion-based model, we use \refbhb\ \citep{refbhb, reisgx}.
Unlike its counterpart, \reflionx, which describes reflection from
supermassive BHs in active galactic nuclei (AGN), \refbhb\ accounts
for the Compton broadening and other effects unique to the hot, dense
discs around stellar-mass BHs.  The model \refbhb\ includes a
pre-packaged thermal component.  This disc component is a
single-temperature blackbody, and hence is intrinsically narrower than
the standard multi-temperature disc spectrum. We remove this hardwired
blackbody component by the following procedure: We construct a lookup
table of blackbody parameters that match the \refbhb\ inputs, and we
use this table to subtract off the best-fitting 2--20~keV blackbody
spectrum, while pairing the reflection portion of \refbhb\ with the
disc model {\sc kerrbb2}.  Although this approach does not account for
radial variations in the reflection spectrum, it has the virtue of
allowing \refbhb\ to be applied more flexibly to spectra with a strong
disc component.

We henceforth refer to this custom version of \refbhb\ as \refbhbm.
The composite of \refbhbm\ operating jointly with \simplr\ and
\kerrbbtwo\ comprises what we refer to hereafter as Model~1.  We also
consider an alternate `prime' version of Model~1 -- Model~1p -- with
\refbhbm\ replaced by its AGN counterpart \reflionx.  Apart from this
substitution, Model~1 and Model~1p are identical.

For the reflection components of each of the models considered here,
we assume Solar metallicity, which is the only setting available for
\refbhb\ and \refbhbm.  The LMC as a whole is known to have a lower
metallicity; however, \lmc\ is likely to be relatively metal rich
because of its young age ($\sim$5 Myr; \citealt{Orosz_2009}).

Our second model family uses \ireflect\ which self-consistently
computes the reflection edges for ionised gas, given an arbitrary
input spectrum. \ireflect\ is a generalisation based on the model {\sc
  pexriv}, but whereas {\sc pexriv} is restricted to a pure power law
for coronal emission, \ireflect\ is freely combined with any coronal
spectrum desired.  As input to {\sc ireflect}, we use the Compton
component generated in the accretion-disc corona, which (as above) is
generated by \simplr\ acting on the disc component \kerrbbtwo.  We
hereafter refer to this composite model as Model~2.  Although
\ireflect\ offers the advantage of computing reflection
self-consistently given an arbitrary input spectrum for the
illuminating hard X-rays, it has several major drawbacks compared to
\refbhb/\refbhbm.  For example, \ireflect\ only computes edge
absorption; that is, it neglects all fluorescent line emission,
including the Fe line itself.  We insert the ionised Fe \ka\ line
emission by adding an intrinsically narrow Gaussian.  Like \refbhbm,
\ireflect\ does not account for radial variation in the disc's
structure.

In summary, we face a tradeoff: Model~1 provides an optimal
description of the atomic features, both lines and edges, whereas
Model 2 provides a superior description of the `continuum' shape of
the reflection component, especially at low energies.  More
specifically, although these two models describe the same process,
Model~2 consistently describes the continuum shape determined jointly
by the thermal, Compton, and reflection components, while fluorescent
line emission is not incorporated self consistently.  Conversely, in
Models~1 and 1p the atomic physics is self consistent, i.e., the
treatment of emission and absorption is unified, but the shape and
intensity of the reflection component is not directly tied to the flux
or curvature in the associated power-law component.  Lacking an
ultimate model of reflection in which the virtues of both models are
captured, we employ both approaches separately in arriving at our
final result.

For all formulations (Model~1, Model~1p, and Model~2), the inner-disc
reflection is convolved with the relativistic smearing kernel
\relconvf\ \citep{relline, Fabian_2012}.  We also include an unblurred
reflection component to account for reflection far from the BH, which
we demonstrate is produced by fluorescence of the stellar wind by the
X-ray source (Section~\ref{subsec:disc:distant}).  We model this sharp
reflection component using \reflionx.

Photoelectric absorption is treated using {\sc tbvarabs}
\citep{Wilms_2000}\footnote{The ISM composition in {\sc tbvarabs} is
  taken from \citet{Hanke_2009}.}.  Our three model formulations in {\sc xspec} notation are: \\
(1)  ~\tbvarabs$\times$(\simplr$\otimes$\kerrbbtwo+\relconvf$\otimes$\\ \qquad \refbhbm+\reflionX), \\
(1p) \tbvarabs$\times$(\simplr$\otimes$\kerrbbtwo+\relconvf$\otimes$\\ \qquad \reflionX+\reflionX), \\
(2) ~\tbvarabs$\times($\simplr$\otimes$\kerrbbtwo+\relconvf$\otimes$\\ \qquad [\ireflect(\simplc)+{\sc gauss}$_{\rm Fe}$]+\reflionX),\\
where \simplc\ is shorthand for the power-law emission isolated from
\simplr$\otimes$\kerrbbtwo.
 
\subsection{Method}\label{subsec:method}

Our results are obtained by first achieving a set of preliminary
spectral fits using {\sc xspec} (v12.7; \citealt{XSPEC}).  These fits
provide the seed for a more intensive and robust analysis using a
Markov chain Monte Carlo (MCMC) routine, which is implemented via a
package described in \citet{Steiner_2011}.  For this application, the
MCMC routine has been modified to work with {\sc xspec}.  For each
fit, a total of 3$\times10^5$ elements are generated\footnote{An
  additional $\approx10^5$ elements were generated during training and
  burn-in phases but these were not used in the final analysis (see
  \citealt{Steiner_2011}).}, which is sufficient to reach
convergence\footnote{In all cases, a satisfactory \citet{Gelman_Rubin}
  convergence diagnostic of $\hat{R} < 1.2$ was obtained.}.  The
results so obtained have been verified using {\sc xspec} in
conjunction with a second MCMC sampler: {\sc emcee} \citep{emcee}.


All but four free parameters were modelled with noninformative priors,
either flat -- for shape parameters such as $\Gamma$ or $T$, or
log-flat (i.e., a flat weighting on the log of the parameter) for
scale parameters such as $\dot{M}$ and component normalisations.  Two
exceptions to this rule are $\nh$, which was assigned a normal prior
distribution\footnote{$N(\mu,\sigma$) refers to a normal distribution
  centred upon $\mu$ with a variance of $\sigma^2$.} of $N(1.15,0.15)
\times 10^{22} \cm^{-2}$ \citep{Hanke_2009}, and the PIN instrumental
background normalisation, which was taken to be a normal distribution
of $N(1,0.03)$ bounded within $1\pm0.06$.  Lastly, based on the
results of \citet{Fabian_2012}, we impose a restriction on the paired
values of $\spin$ and the inner emissivity index $q_1$.  In
\citet{Fabian_2012}, it was shown that for all non-maximal values of
spin ($\spin \lesssim 0.94$), $q_1$ is not expected to deviate
substantially from its nominal thin-disc value of $q=3$.  High or
extreme values of $q_1$ are expected only for strong distortions of
spacetime near the horizon, for $R \lesssim 2 GM /c^2$ (e.g.,
\citealt{Wilkins_2011}).  To prevent runaway to unreasonable regions
of parameter space, we assign a weakly constraining prior on $q_1$, a
normal distribution of the following form:
\begin{equation}
p(q_1|\spin)= N(3, 0.475+0.05\times(1-\spin)^{-1}).
\end{equation}
Under this prior, at the minimum possible spin, $\spin=-1$, the
prior's standard deviation is 0.5, and it increases slowly with spin,
reaching unity at $\spin\approx0.9$.  Thus, for most allowed values of
spin, the fit is loosely constrained to conform to its nominal value,
$q_1=3$.  At the highest values of spin, the prior offers essentially
no constraint because the standard deviation diverges.

\section{Analysis}\label{section:results}

Before applying these spectral models in detail, we first step back
and adopt a more data-centric approach that firmly establishes the
existence of a relativistically broadened Fe line.  LMC X--1 has a
sibling BH binary in the Large Magellanic Cloud, LMC X--3.  LMC X--3
is also persistently bright, and it is nearly always in a soft state.
However, in contrast to LMC X--1, LMC X--3 is highly variable with
$\lledd$ ranging from $\lesssim1$ per cent to $\sim50$ per cent
\citep{Wilms_2001, Steiner_2010}.  Moreover, the spectrum of LMC X--3
appears featureless, to such extent that it was recently used by
\citet{Kubota_2010} as a benchmark for testing the performance of
spectral disc models, which makes it an ideal reference source for our
purposes.  Use the $\sim 70$~ks \suzaku\ spectrum of LMC X--3 obtained
in December 2008, studied by Kubota et al., and following precisely
the same reduction procedure we used for LMC X--1, we compute the
ratio of the spectra of the two sources for the combined
front-illuminated XIS units.  The virtue of this approach is that this
ratio spectrum is completely free of any detector-calibration or
data-reduction artefacts.

\begin{figure}
{\includegraphics[clip=true, angle=90,width=8.8cm]{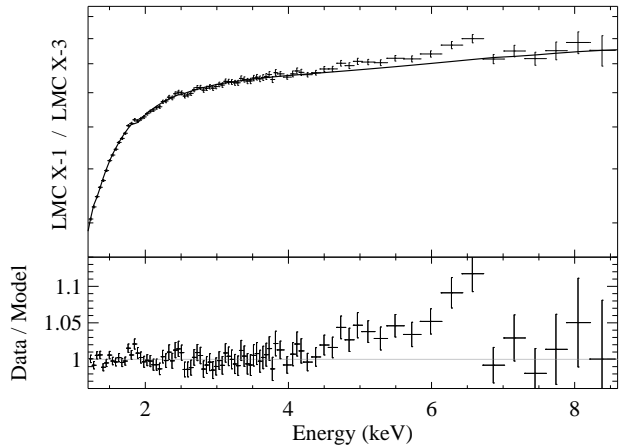}} 
\caption{Top: Ratio of the spectrum of \lmc\ to that of LMC X--3 shown
  fitted to a polynomial (solid line).  Bottom: The residual spectrum,
  which clearly reveals the presence of broadened Fe \ka\ emission.}
\label{fig:ratio}
\end{figure}

Because the two continua have different shapes, the ratio spectrum has
some smooth curvature, which we remove by fitting it to a low-order
polynomial.  The residuals clearly reveal the presence of a broad
Fe-line component in LMC X--1, as shown in Fig.~\ref{fig:ratio}.  The
line is peaked near 6.7 keV, and it has a broad red wing, as expected
for a gravitationally-redshifted line fluoresced in the inner disc.

The broad component extends to below 5 keV at an inclination of
$i=36\degr$ \citep{Orosz_2009}, which implies that the disc inner
radius is $r < 2.5 GM/c^2$ and the spin is correspondingly high.
While one cannot obtain a definitive measurement of spin using this
ratio spectrum alone, it nevertheless serves to confirm the presence
of LMC X--1's broad Fe line and suggests that the BH is rapidly
spinning.

\begin{figure}
{\includegraphics[clip=true, angle=90,width=8.8cm]{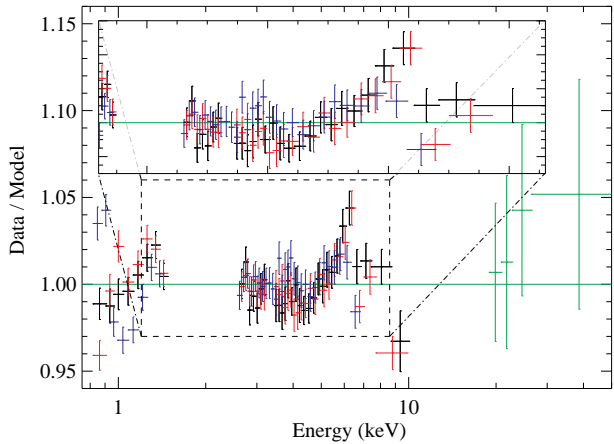}}
\caption{Residuals to the best fit of a model with Comptonised disc
  emission and unblurred (narrow) reflection reveal a prominent broad
  red wing to the Fe line in the ratio spectrum.  The PIN data show a
  hint of curvature in the energy range of the Compton hump.  These
  features further indicate the presence of relativistic reflection
  from the inner disc.  Here, the data have been rebinned for plotting
  purposes only. To allow for a straightforward comparison, the inset
  highlights the same energy range as Fig.~\ref{fig:ratio}.
}\label{fig:ratiofit}
\end{figure}

As a first fit to the spectral data, without yet treating effect of
relativistic reflection, we apply a model of Comptonised disc emission
combined with a reflection component which is unblurred and narrow
(\tbvarabs$\times$(\simplr$\otimes$\kerrbbtwo+\reflionX)).  The
data-to-model ratio for the best fit ($\rchinu ~ 1.2$) is shown with
Fig.~\ref{fig:ratiofit}, in which the residual features from 4--6~keV
highlight an excess attributed to an unmodelled broad red wing of the
Fe line.  Likewise, the deficit above 8 keV and the excess and slight
curvature above 20 keV indicate, respectively, the smeared Fe edge and
Compton hump (although the Compton hump signal is statistically marginal).  These residual features are indicators of an untreated
component of relativistically broadened reflection.

Having established the existence of gravitationally-broadened spectral
reflection features, we will now apply the suite of disc/reflection
spectral models described above and thereby constrain LMC X--1's spin.

\subsection{Spectral Analysis}

We begin by making the standard assumption of alignment between the BH
and the binary orbital plane ($i \approx 36\degr$;
\citealt{Orosz_2009}), and we proceed to apply Models~1, 1p, and 2 to
the \suzaku\ data alone.  We then follow by applying these same models
to the combined \suzaku\ and \rxte\ data set.  The dominance of the
thermal component in the spectrum of LMC X--1, and the modest
luminosity of the source (Section~\ref{section:Intro}), assure that
the spectrum is well described by standard thin-disc theory.
Therefore, in modelling the inner reflection component, we begin by
assuming that illumination of the disc follows a thin-disc profile
with the standard value of the emissivity index, $q=3$
(\citealt{Fabian_1989}).

\begin{figure}
{\includegraphics[clip=true, angle=90,width=8.9cm]{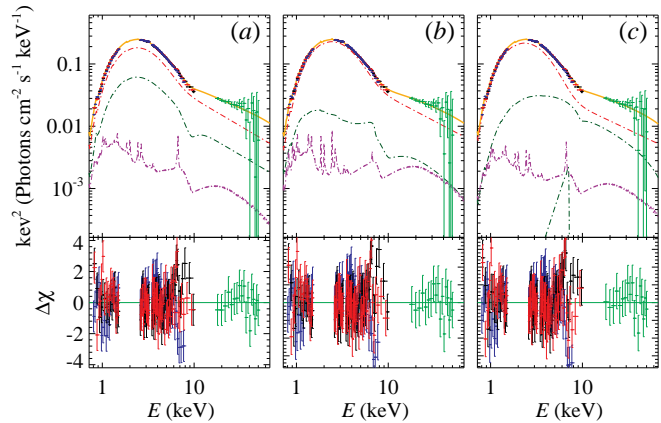}}
\caption{The unfolded spectral fit and fit residuals for LMC X--1
  using all three spectral models with $q=3$. We show the fit for
  Model~1 (\refbhbm), our adopted model, in panel $a$; Model~1p
  (\reflionX) in panel $b$; and Model~2 (\ireflect) in panel $c$.  The
  reflection signal is comprised of two parts: The primary comes from
  the inner portion of the BH's accretion disc (green lines), and the
  secondary from the distant gas in the system (purple lines).
  Comptonised disc emission is shown in red, and the composite model
  is overlaid in orange.  In panel $a$, because \refbhbm\ operates as
  reflection/reprocessing modifying a thermal signal, the relativistic
  reflection is drawn showing
  \refbhbm+$1.5\fsc($\simplr$\otimes$\kerrbbtwo$)$, arbitrarily chosen
  in order to depict a comparable normalisation to panels $b$ and $c$.
  The XIS spectra have been binned to a minimum signal-to-noise of 70
  for plotting only.}\label{fig:bestfit}
\end{figure}

Fig.~\ref{fig:bestfit} shows our best fits to the \suzaku\ spectra for
all three models.  The fitting results are summarised in
Table~\ref{tab:results} ($q=3$ results are denoted by subtypes (i) and
(iii)).  MCMC is designed to directly provide posterior probability
densities for the free parameters of our models.  However, for large
numbers of free parameters MCMC alone does not provide an optimised
estimate of the minimum value of $\rchinu$.  Therefore, after running
the MCMC chains, in order to provide an estimate of the usual goodness
of fit, we have optimised the model $\rchinu$ about the central values
in the chain; these optimised fits yield the $\rchinu$ values given in
Table~\ref{tab:results}.  Although the MCMC chains do not directly
deliver optimised goodness-of-fit estimates, we emphasise that the
chains themselves provide the most direct estimates of the probability
distributions for our model parameters, which is of chief interest.

\begin{figure}
{\includegraphics[clip=true, angle=90,width=8.8cm]{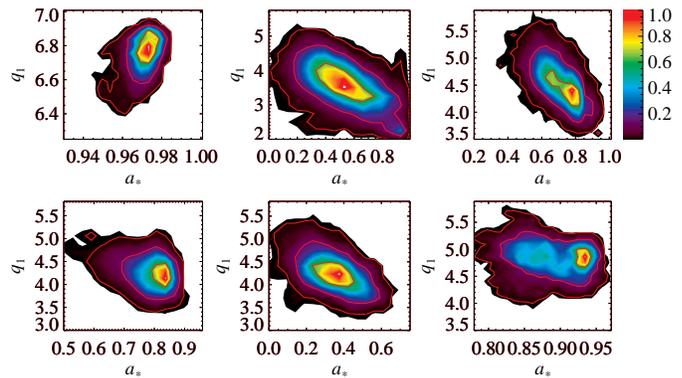}}
\caption{Selected correlation results from the MCMC analysis showing
  that spin and the inner emissivity index $q_1$ are weakly
  correlated.  Red contours show 1$\sigma$, 2$\sigma$, and 3$\sigma$
  confidence limits about maximum density values.  Top panels
  correspond to \suzaku\ fitted alone (model subtype (ii)), and the
  bottom panels show results for \rxte\ and \suzaku\ fitted together
  (model subtype (iv)).  Panels on the far left use \refbhbm\
  (Model~1), those in the centre column use \reflionX\ (Model~1p), and
  those on the right use \ireflect\ (Model~2).}\label{fig:spinlimits}
\end{figure}

We explore the effect of allowing $q$ to vary and to take on the form
of a broken power law (model subtypes (ii) and (iv) in
Table~\ref{tab:results}).  For moderate or low spins, such as those
given for Model~1p, this exercise is not physically motivated, but we
nevertheless use this approach to understand the effect of having
initially fixed $q$ at its canonical value of 3.  Once $q_1$ is freed,
all three models return high values, $q_1>3$; the outer index is fixed
to $q_2=3$.  As a net result of this exercise, the values of spin are
slightly depressed.  The dependence of $\spin$ on $q_1$ is illustrated
in Fig.~\ref{fig:spinlimits}.  There is generally a weak
anti-correlation between the two parameters, which is most pronounced
for Models~1p (center) and 2 (right).

Because a high value of the inner emissivity index $q_1$ indicates
substantial light bending in the strong-gravity environment close to
the black hole's event horizon, one expects high $q_1$ to be
associated with high values of spin.  That we find the opposite
correlation in our fits -- high $q_1$ is associated with low spin --
is simply a systematic artefact of either our model or the data, which
is possibly related to the limited signal-to-noise in the reflection
features.

As is evident in Fig.~\ref{fig:bestfit}, the spin is obtained by
decomposing the blurred and sharp reflection components, with a
substantial fraction of this signal coming from around the Fe line and
edge.  The sharp reflection is produced by material distant from the
source, which we identify with the wind of the companion star.  This
subject is discussed in Section~\ref{subsec:disc:distant}.

Overall, the three models perform comparably well.  (The statistical
differences in the quality of fit between Models 1, 1p, and 2 are even
less when one factors in the influence of the priors -- an effect not
captured by $\rchinu$.)  However, because Model~1p uses a reflection
model designed for AGN, whereas Model~1 is optimised for the densities
and environments of BH binaries, we favour Model~1 because it is
physically more appropriate.  Model~2, which offers the most
self-consistent treatment of the broad reflection continuum (while
ignoring the fluorescent emission features), also returns a high spin.
Model~1p is less constraining, but consistent with the others.
Although we adopt the results for Model~1 which we consider to be the
most physical of the three models, we use the combined results for all
the models in all the columns of Table~\ref{tab:results} to account
for systematic uncertainty in our spin measurement.  In particular,
although their paired fits for $q_1$ and $\spin$ are physically
suspect (described above), we nevertheless include Model~1p(ii),
Model~2(ii), and Model~1p(iv) in determining the net result.  This
conservative approach increases our final uncertainty.

\begin{figure}
{\includegraphics[clip=true, angle=90,width=8.8cm]{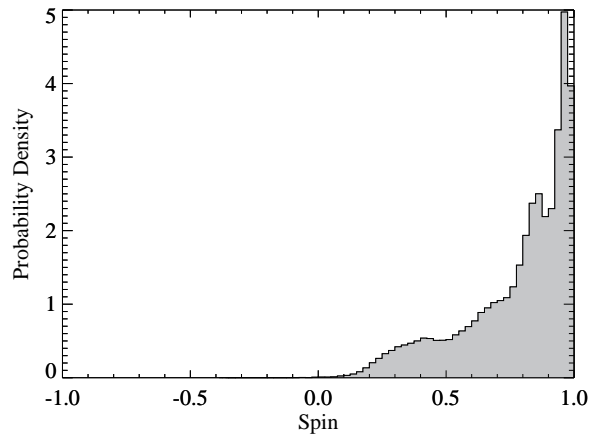}}
\caption{Our final and adopted spin probability distribution obtained
  by combining results from all the models in Table~\ref{tab:results}
  with equal weights.  }
\label{fig:probability}
\end{figure}

Our final result for Model~1, combining all four variants (i.e.,
(i)-(iv) in Table~\ref{tab:results}) is $\spin =
0.97^{+0.01}_{-0.13}$.  We account for systematic effects by combining
the results for all the models in all the columns of
Table~\ref{tab:results} (with equal weight assigned to each), and
thereby obtain $\spin = 0.97^{+0.02}_{-0.25}$.  As a firm lower bound,
at 99 per cent confidence, the spin is constrained $\spin > 0.2$.
This distribution, computed from the sum of all the MCMC chains, is
shown in in Fig.~\ref{fig:probability}.

In order to assess the effect of assuming Solar abundances, we explore
a reduced metallicity of $Z=0.3 Z_\ssun$ for Model~1p (this
metallicity setting is possible using \reflionx\ but not \refbhbm).
We find an increase of $\lesssim 10$ per cent in the ISCO radius,
which implies a reduction in the spin of $\Delta\spin \approx 0.02$
from our fiducial values.  Metallicity is, therefore, a relatively
minor source of systematic uncertainty.  Similarly, we have explored
the uncertainty introduced by the 1.9 deg error in the orbital
inclination \citep{Orosz_2009} and find that the resulting spin
uncertainty is quite small ($\Delta\spin \lesssim 0.01$) compared to
our net uncertainty.  Lastly, one notable deficiency of Model~2 is
that it omits treatment of Compton broadening of the Fe line due to
scattering in the (keV-temperature) disc atmosphere.  Because this
temperature is far below the energy of the Fe line, Compton broadening
produces an intrinsic red tail to the line (e.g.,
\citealt{Torrejon_2010, Watanabe_2003}).  We explore this broadening
by replacing the narrow Gaussian of Model~2 with a one-sided Gaussian
with width $\sigma = 0.5$~keV.  The width used is approximately twice
as large as that due to single scattering in the disc.  The broadened
line returns significantly improved fits ($\Delta\chisq > 15$), which
we interpret as a likely detection of some degree of Compton
broadening.  In all instances of Model~2 (i.e., (i)-(iv) in
Table~\ref{tab:results}), the spin returned is in the range $\spin =
0.94 - 0.98$.  We conclude that, because Compton broadening is an
order of magnitude smaller than the maximum gravitational redshift for
a rapidly-spinning source, the impact on the fitted spin is moderate.
Generally, accounting for this extra broadening causes the model to
return a slightly lower value for spin in Model~2.

\section{Discussion}\label{section:discussion}

\subsection{The Connection Between the Compton Power Law and Reflection}\label{subsec:disc:corona}

\begin{figure}
{\includegraphics[clip=true, angle=90,width=8.8cm]{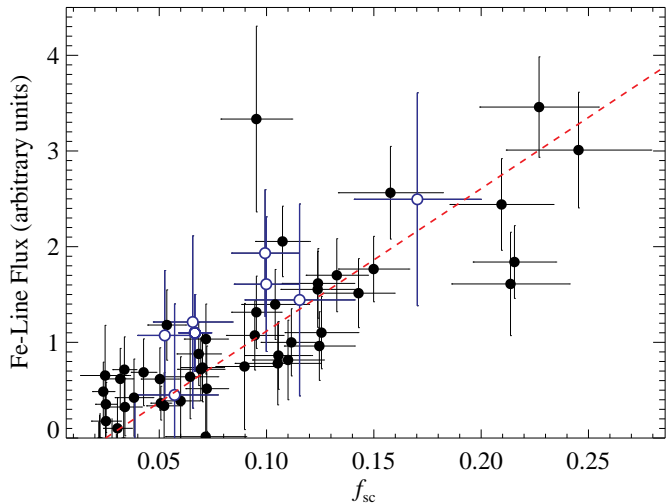}}
\caption{A strong correlation exists between the relative strength of
  the Compton power-law component ($\fsc$) and the Fe-line flux based
  on \rxte\ data.  Because the simplistic model employed assumes an
  arbitrary line width, the line flux is shown to arbitrary scale.
  The \citet{Gou_2009} spectra are displayed as closed symbols, and
  our observations are plotted as open symbols.  A linear fit is
  overlaid as a dashed line.}\label{fig:fsc_fe}
\end{figure}

We now explore the relationship between the most prominent feature of
the reflection component, namely the Fe line, and the degree of
Comptonisation.  To do this empirically, we make use of a simplified
continuum model employed in \citet{Gou_2009}, with a small
modification: \tbvarabs(\simpl$\otimes$\kerrbbtwo+{\sc gauss}$_{\rm
  Fe}$)\footnote{For all spectra, we use fixed values for the Gaussian
  line energy and width (6.5~keV and 0.25~keV respectively), and we
  use the continuum-fitting value of spin obtained by Gou et
  al. ($\spin=0.92$).  Further following \citet{Gou_2009}, we restrict
  the power-law index to values between $\Gamma=2.25$ and
  $\Gamma=3.25$.}.  We apply this model to both the full
\citeauthor{Gou_2009} data sample of 53 \rxte\ spectra and to the
\rxte\ spectra collected during our \suzaku\ observation.  This model
lacks the physical rigour of those we have been considering, but it
has the virtue of being sensitive to the {\em strengths} of the Fe
line and Compton component without requiring additional
parameterisation\footnote{Because this simplistic model does not
  account for photoabsorption of Compton-backscattered photons, which
  return to the disc, the value of $\fsc$ is lower here by a factor of
  roughly two compared to its value under Models~1, 1p, and 2 above.}.

As shown in Fig.~\ref{fig:fsc_fe}, a clear and positive correlation is
observed which is well-described by a linear fit ($\rchinu = 1.17$)
with a correlation that is significant at $>12\sigma$.  Regardless of
the geometry of the corona -- whether a disc-hugging skin, a
lamp-post, or a centrally-concentrated cloud -- the existence of this
correlation confirms that the coronal Compton emission is the primary
source that illuminates the disc and produces the
reflection/fluorescence component in LMC X--1.  This result indicates
that disc self-irradiation (which could be important in producing
reflection features either at high luminosities or in Compton-weak
sources) is not dominant here.  Our findings are in close agreement
with those of \citet{Hiemstra_2011}, who demonstrate for the BH XTE
J1652--453 the coexistence of a bright disc and strong reflection
features.  Hiemstra et al. point out that strong coronal flows in
bright (intermediate-state) discs can produce the most prominent broad
Fe lines, which may explain the strong line feature we observe in LMC
X--1.

\subsection{The Thermal Continuum}\label{subsec:disc:CF}

It is often problematic to use the same spectrum to estimate spin by
both the continuum-fitting and Fe-line methods because the optimal
spectrum for continuum-fitting has a weak power-law/reflection
component, whereas a spectrum that is optimal for the reflection
method generally has a dominant power-law component, which compromises
continuum-fitting measurements.  Unfortunately, for the \suzaku\ and
\rxte\ spectra treated here, one cannot obtain a reliable spin
measurement via continuum fitting.  Despite its stable luminosity, LMC
X--1 exhibits strong rms power fluctuations\footnote{We follow the
  convention of \citet{RM06}, who define rms variability as the
  average power from 0.1--10 Hz in the band 2--20~keV.}.  To obtain
reliable continuum-fitting results, one generally restricts the rms
power to be $\lesssim 8$ per cent (e.g., \citealt{Gou_2009, RM06}).
However, for the spectrum in question, the rms power is in the range
27--35 per cent, and therefore a reliable continuum-fitting spin
result cannot be obtained from these data.

In a study of LMC X--1's variable power-law component,
\citet{Ruhlen_2011} examine the archive of \rxte\ observations of LMC
X--1, and, assuming a uniform and nominal power-law index, find a weak
anticorrelation between disc flux and temperature.  This is exactly
analogous to the anticorrelation of spin and $\dot{M}$ shown in
figure~4 of \citet{Gou_2009}.  Ruhlen et al. interpret their flux,
temperature anticorrelation as implying the presence of a variable
corona that sometimes obscures the inner disc.  While this
interpretation is possible, it implies the occasional presence of an
optically-thick, $\sim 50$ keV coronal cloud, which has never been
observed.  We therefore consider this explanation unlikely.  This
anticorrelation may instead be related to instabilities in the disc,
as suggested by the correlation between the strength of the intense
power-law component and the rms variability in the power spectrum
(\citealt{Gou_2009}; and with the reflection strength, see
Section~\ref{subsec:disc:corona}).  Speculatively, the variability
noted by \citeauthor{Ruhlen_2011} may be consistent with the
inhomogeneous-disc model of \citet{Dexter_2012}.

For any of these interpretations, the \citet{Gou_2009}
continuum-fitting spin measurement is robust to these effects because
for measuring spin, Gou et al. selected only those data in which the
Compton component is weakest ($\fsc < 0.075$), or equivalently, data
for which the rms variability is weak.  However, the continuum-fitting
measurement is confounded for the \suzaku\ observation considered
here.

We note that the sign and magnitude of the deviation we find for the
continuum-fitting value of spin compared to the result of
\citet{Gou_2009} is consistent with what has been obtained for other
systems; namely at excessively large values of rms, the
continuum-fitting model tends to return erroneously small values of
spin (e.g., fig.~1 in \citealt{Steiner_2009}).

Because in this instance the thermal continuum model is flawed, and
additionally because \citet{Kubota_2010} emphasise that a true disc
spectrum may be intrinsically broader than the zero-torque \kerrbb\
model, we examine the effect of using a modified torque at the
inner-boundary.  For the reasonable values we have explored, $\eta <
10$ per cent (see \citealt{KERRBB} for details on the torque
prescription), the reflection spin measurement is insignificantly
affected by this change in the continuum model.  To otherwise assess
the impact of our choice of thermal continuum model, we have tested
replacing \kerrbbtwo\ with \diskbb\ \citep{DISKBB}, and also with \bhspec\
\citep{BHSPEC} in Models~1(i), 1p(i), and 2(i).  The goodness-of-fits
returned with these disc models is worse by $\Delta\chisq \approx 20$,
and the spin in each instance is consistent with $\spin > 0.9$.  We
conclude that our reflection spin result is robust to the choice of
thermal continuum model.

\subsection{Distant Reflection}\label{subsec:disc:distant}

\begin{figure}
{\includegraphics[clip=true, angle=90,width=8.8cm]{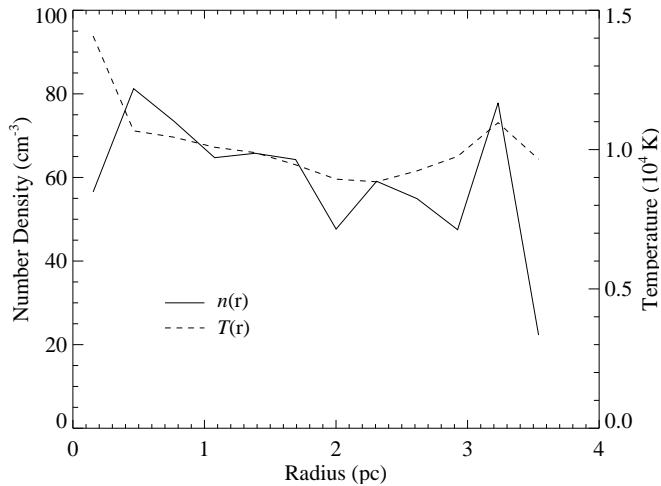}}
\caption{Radial profiles for the density and gas temperature in the
  ionisation cone are approximately flat.  This supports our assertion
  that the optical ionisation cone is a shell of shocked interstellar
  gas.  This gas encloses a bubble inflated by the stellar wind (e.g.,
  compare with the `shell' region in fig.~3 of
  \citealt{Weaver_1977}).}
\label{fig:ryan}
\end{figure}

Based on optical integral-field spectroscopy with a resolution of
$\sim1$ pc, \citet{Cooke_2008} mapped a large ($\approx 4$~pc)
cone-shaped ionisation nebula enshrouding LMC X--1
\citep{Pakull_Angebault_1986}.  The cone has an opening angle
$\sim45\degr$ and a mean density on large scales of
$n\lesssim100~\cm^{-3}$.  Meanwhile, our spectral fits imply the
presence of a strong source of reflection that is distant from LMC
X--1 (i.e., separated by $r >> 10^8~\cm$) with a luminosity $\sim
10^{36} - 10^{37}$~erg~s$^{-1}$.

Both the cone structure and the reflection component are readily
explained at once by positing the presence of a powerful wind that is
photoionised by the luminous X-ray source.  We show that the wind is
likely supplied by a dominant outflow from the O supergiant companion
\citep{Orosz_2009} (possibly coupled with a fast wind launched from
the disc).  The source of illumination, meanwhile, is the
self-irradiated accretion disc, which supplies the requisite flux of
ionising photons after reprocessing $\lesssim 5$ per cent of the
X-rays from the inner disc into ionising photons in the outer disc
(radii $\gtrsim 100 GM/c^2$; \citealt{DISKIR, Gou_2009}).

Adopting a very simplified model in which reflection is approximated
as efficient Thomson scattering through an idealised wind with
ionisation parameter $\sim 10^3$ erg\ cm\ s$^{-1}$, we find that the
wind is launched at a distance $r_0 \sim 0.2$~AU from the BH and with
a density of $n_0 \sim 10^{10}$~cm$^{-3}$.  These values match the
BH-to-companion star distance and the surface density in its stellar
wind (\citealt{Lamers_1993, Orosz_2009}); this identifies the stellar
wind as the likely source of distant ionised gas.  At sub-pc scales,
the ionisation of the wind is constant ($n \propto r^{-2}$), but
because density falls off rapidly, most of the (sharp) reflection is
centrally concentrated and emitted from the innermost several AU.

Meanwhile, at parsec scales, the wind is shocked against the ambient
interstellar medium (ISM).  As a result, the ionisation cone is mostly
filled with a dense gas of shocked wind.  This cone, in turn, is
enclosed in a thin shell of even denser shocked ISM (e.g.,
\citealt{Weaver_1977}).  This picture is borne out by the flat
temperature and density profiles\footnote{These profiles are derived
  from the data in \citet{Cooke_2008} by using the [S II]
  $\lambda\lambda$6716/6731 ratio as a density diagnostic and the [O
  III] $\lambda\lambda$(4959+5007)/4363 ratio as a temperature
  diagnostic \citep{Osterbrock_book}. See \citet{Cooke_2008} for more
  details.}  within the ionisation cone, which are shown in
Fig.~\ref{fig:ryan}.

Adopting this simplistic description of the wind as valid, and
employing the \citet{Weaver_1977} wind-bubble model, the projected
mass in the wind is of order $\sim 0.1 \msun$, which suggests a
time-scale of $\sim 10^6$ years to produce the ionisation cone
surrounding LMC X--1.\footnote{Derived using the ISM density in LMC
  X--1's local environment (an O-B association) of $\sim
  20$~cm$^{-3}$.  This, in turn, is obtained by assuming the standard
  jump condition at the outer shock boundary given that the density in
  Fig.~\ref{fig:ryan} corresponds to the high-density outer shell.}
The mass in the cone is then commensurate with the integrated mass
loss from the companion star, and a time-scale results which is
comparable to the $\approx5$~Myr age of the system
(\citealt{Orosz_2009}).

\section{Conclusions}\label{section:conc}

We have analysed a deep \suzaku\ and \rxte\ observation of LMC X--1,
and we have demonstrated that a broad Fe \ka\ line is present in the
spectrum of the source.  Using existing and new spectral reflection
models, we have measured the BH's spin parameter to be $\spin =
0.97^{+0.01}_{-0.13}$ for our favoured model (Model~1), and to be
$\spin = 0.97^{+0.02}_{-0.25}$ when making an allowance for systematic
error by considering our full range of models.  At 99\% confidence, we
establish $\spin > 0.2$. Both of these spin estimates are in agreement
with the spin determined using the X-ray continuum-fitting method
\citep{Gou_2009}.

Apart from the measurement of spin, we present two additional results:
(1) For a large sample of \rxte\ spectra, we demonstrate a strong and
positive correlation between the Compton and reflection components.
(2) Far from the strong gravity environment of the BH, we identify the
wind of the massive companion star as the source of a luminous and
sharp reflection component. We conclude that the wind and persistent
X-ray source together maintain the parsec-scale ionisation cone that
envelops the binary system.

\section*{Acknowledgments}

It is a pleasure to thank Shin'ya Yamada, Jon Miller, Aya Kubota, and
Kazuo Makishima for their input.  J.F.S. thanks the \rxte\ team for
their fast and helpful assistance with the TOO observation, and
particularly Evan Smith of the \rxte\ team and Koji Mukai of \suzaku\
for their ready advice on conducting the observations.  We also thank
the anonymous referee for a helpful report which improved this work.
J.F.S. was partially supported by the Smithsonian Institution
Endowment Funds.  R.C.R. is supported by NASA through the Einstein
Fellowship Program, grant No.  PF1-120087 and is a member of the
Michigan Society of Fellows.  J.E.M. acknowledges
support from NASA grants NNX11AD08G and NNX09AV59G.\\

{\it Facilities:} \suzaku, \rxte

\newcounter{BIBcounter}        
\refstepcounter{BIBcounter}

\bibliography{js}

\clearpage
\onecolumn
  \begin{deluxetable}{lcccccc}
  \tabletypesize{\scriptsize}
   \setlength{\tabcolsep}{3pt}
  \tablecolumns{7}
  \tablewidth{0pc}
  \tablecaption{Spectral Models}
  \tablehead{ Parameters    & Model~1(i)      & Model~1p(i)      & Model~2(i)  &  Model~1(ii)  & Model~1p(ii)   & Model~2(ii)  \vspace{0.4mm}\\
              Instruments   & \suzaku       & \suzaku       & \suzaku   &  \suzaku   & \suzaku    & \suzaku   }
  \startdata
$N_{H}  (10^{22} \cm^{-2}) $                                           &  $                1.37 \pm 0.01 $  &  $         1.54^{+ 0.02}_{- 0.03} $  &  $      1.38^{+ 0.02}_{- 0.01} $  &  $      1.29^{+ 0.02}_{- 0.01} $  &  $                  1.55 \pm 0.03 $  &  $    1.41^{+ 0.01}_{- 0.03} $   \\ 
$\Gamma$                                                               &  $         2.62^{+ 0.10}_{- 0.06} $  &  $                  2.73 \pm 0.06 $  &  $                  2.65 \pm 0.04 $  &  $         2.50^{+ 0.10}_{- 0.05} $  &  $                  2.74 \pm 0.06 $  &  $                2.62 \pm 0.05 $   \\ 
$\fsc$\tablenotemark{a}                                                &  $         0.17^{+ 0.03}_{- 0.01} $  &  $                0.204 \pm 0.014 $  &  $      0.13 \pm 0.01 $  &  $                0.121 \pm 0.017 $  &  $      0.20^{+ 0.02}_{- 0.01} $  &  $    0.13 \pm 0.01 $   \\ 
$a_{*, {\rm CF}}$\tablenotemark{b}                                     &  $       0.86^{+ 0.01}_{- 0.02} $  &  $                  0.81\pm0.01 $  &  $                0.82 \pm 0.01 $  &  $      0.927^{+ 0.005}_{- 0.011} $  &  $          0.81 \pm 0.01 $  &  $              0.82 \pm 0.01 $   \\ 
$\Mdot$\tablenotemark{b}$ (10^{18}{\rm g~s}^{-1})$                     &  $         1.88^{+ 0.05}_{- 0.03} $  &  $         1.77^{+ 0.04}_{- 0.05} $  &  $                1.63 \pm 0.02 $  &  $         1.66^{+ 0.16}_{- 0.06} $  &  $                  1.78 \pm 0.04 $  &  $                1.63 \pm 0.02 $   \\ 
$q_1$\tablenotemark{c}                                                 &                                    3 &                                    3 &                                    3 &  $         6.78^{+ 0.12}_{- 0.06} $  &  $                    3.7 \pm 0.5 $  &  $          4.5^{+ 0.4}_{- 0.2} $   \\ 
$R_{\rm br} (R_{\rm ISCO})$\tablenotemark{c}                           &                            \nodata   &                            \nodata   &                            \nodata   &  $              2.99 \pm 0.04 $  &  $         4.9^{+ 0.1}_{- 2.} $  &  $       4.9^{+ 0.1}_{- 0.7} $   \\ 
$a_{*, {\rm ref}}$                                                     &  $         0.94^{+ 0.02}_{- 0.20} $  &  $       0.99^{+ 0.01}_{- 0.31} $  &  $           0.99^{+ 0.01}_{- 0.11} $             &  $      0.97\pm0.01 $  &  $               0.54 \pm 0.19 $  &  $     0.82^{+ 0.01}_{- 0.18} $   \\ 
$H_{\rm den}$ (10$^{19}$~cm$^{-3}$)                                    &  $              15^{+ 2}_{- 9} $  &                            \nodata   &                            \nodata   &  $            5.0^{+ 2.3}_{- 0.7} $  &                            \nodata   &                          \nodata    \\ 
$kT_{\rm ref}$ (keV)                                                   &  $         0.73^{+ 0.01}_{- 0.02} $  &                            \nodata   &                            \nodata   &  $      0.85^{+ 0.01}_{- 0.02} $  &                            \nodata   &                          \nodata    \\ 
 Illum/BB                                                              &  $            0.0204 \pm 0.0007 $  &                            \nodata   &                            \nodata   &  $   0.0125^{+ 0.002}_{- 0.001} $  &                            \nodata   &                          \nodata    \\ 
$N_{\rm ref} (10^{-7}) / (10^{-3})$\tablenotemark{d}                   &  $            1.6^{+ 0.4}_{- 0.3} $  &  $            1.1^{+ 0.3}_{- 0.2} $  &                            \nodata   &  $            7.8^{+ 1.1}_{- 1.7} $  &  $            1.2^{+ 0.3}_{- 0.2} $  &                          \nodata    \\ 
$\xi$\tablenotemark{e} (erg\ cm\ s$^{-1}$)                             &                            \nodata   &  $        9700^{+ 100}_{- 2500} $  &  $     15000^{+ 1000}_{- 8000} $  &                            \nodata   &  $      9000^{+ 1000}_{- 2000} $  &  $   14000^{+ 2000}_{- 5000} $   \\ 
$T_{\rm ref} $ ($10^6$~K)                                              &                            \nodata   &                            \nodata   &  $               6^{+ 3}_{- 4} $  &                            \nodata   &                            \nodata   &  $            15^{+ 1}_{- 4} $   \\ 
$E_{\rm Fe} $\tablenotemark{f}(keV)                                    &                            \nodata   &                            \nodata   &  $      6.65^{+ 0.02} $  &                            \nodata   &                            \nodata   &  $    6.65^{+ 0.04} $   \\ 
$F_{\rm Fe} $\tablenotemark{f}$(10^{-4}{\rm ph~s}^{-1}\cm^{-2})$       &                            \nodata   &                            \nodata   &  $                    1.1 \pm 0.3 $  &                            \nodata   &                            \nodata   &  $                  2.1 \pm 0.4 $   \\ 
$\xi_{D} $\tablenotemark{e,g}  (erg\ cm\ s$^{-1}$)                     &  $                 3300 \pm 900 $  &  $           200 \pm 100 $  &  $             15^{+ 100}_{- 5} $  &  $       2800^{+ 1600}_{- 600} $  &  $                   230 \pm 90 $  &  $          20^{+ 100}_{- 5} $   \\ 
$N_{\rm ref, D}$\tablenotemark{g}$ (10^{-7})$                          &  $            0.5^{+ 0.3}_{- 0.1} $  &  $            35^{+ 11}_{- 22} $  &  $                    33 \pm 20 $  &  $            0.5^{+ 0.2}_{- 0.1} $  &  $             20^{+ 20}_{- 5} $  &  $           30 \pm 10 $   \\ 
$N_{\rm XIS 3} / N_{\rm XIS 0}$\tablenotemark{h}                       &  $      0.960^{+ 0.001}_{- 0.002} $  &  $      0.960^{+ 0.001}_{- 0.003} $  &  $              0.960 \pm 0.002 $  &  $              0.960 \pm 0.002 $  &  $      0.960^{+ 0.001}_{- 0.003} $  &  $    0.960^{+ 0.001}_{- 0.003} $   \\ 
$N_{\rm XIS 1} / N_{\rm XIS 0}$\tablenotemark{i}                       &  $              0.916 \pm 0.002 $  &  $      0.921^{+ 0.002}_{- 0.004} $  &  $                0.919 \pm 0.004 $  &  $      0.917^{+ 0.004}_{- 0.003} $  &  $                0.919 \pm 0.003 $  &  $              0.918 \pm 0.003 $   \\ 
$\Delta\Gamma(XIS 1 - XIS 0) $\tablenotemark{i}                        &  $             -0.017 \pm 0.002 $  &  $     -0.014^{+ 0.003}_{- 0.002} $  &  $             -0.015 \pm 0.003 $  &  $     -0.014^{+ 0.002}_{- 0.004} $  &  $               -0.014 \pm 0.003 $  &  $   -0.015^{+ 0.002}_{- 0.003} $   \\ 
$N_{\rm PIN~CXB}$\tablenotemark{j}$(10^{-4}{\rm ph~s}^{-1}\cm^{-2})$   &  $            9.3^{+ 0.6}_{- 0.7} $  &  $                    8.8 \pm 0.6 $  &  $                    8.8 \pm 0.7 $  &  $                    9.1 \pm 0.7 $  &  $                    8.9 \pm 0.7 $  &  $                  8.9 \pm 0.7 $   \\ 
PIN$_{\rm BG, Inst}$ Norm\tablenotemark{k}                             &  $       1.03^{+ 0.01}_{- 0.02} $  &  $        1.03^{+ 0.02}_{- 0.01} $  &  $        1.04^{+ 0.01}_{- 0.02} $  &  $        1.03^{+ 0.01}_{- 0.02} $  &  $                 1.04 \pm 0.01 $  &  $              1.03 \pm 0.01 $   \\ 
$N_{\rm PIN}  / N_{\rm XIS 0} $\tablenotemark{h}                       &                                 1.16 &                                 1.16 &                                 1.16 &                                 1.16 &                                 1.16 &                               1.16  \\ 
   \hline                                                                      
       $\chi^2/\nu$                                                    &                 395.4/396            &                     396.3/398        &                415.0/395             &                 389.0/394            &                    375.3/396           &                    387.3/393          \vspace{0.25mm}      \\     
          \hline
          \hline
          Parameters    & Model~1(iii)   & Model~1p(iii)    & Model~2(iii)  & Model~1(iv)  & Model~1p(iv) & Model~2(iv) \vspace{0.5mm} \\  
           Instruments   & \suzaku    & \suzaku     & \suzaku   & \suzaku  & \suzaku  & \suzaku    \\   
           \vspace{0.4mm}              & \& \rxte   & \& \rxte    & \& \rxte  & \& \rxte  & \& \rxte & \& \rxte   \\
           \hline
$N_{H}  (10^{22} \cm^{-2}) $                                           &  $      1.364^{+ 0.011}_{- 0.005} $  &  $                  1.55 \pm 0.02 $  &  $      1.44 \pm 0.01 $  &  $      1.39 \pm 0.01  $  &  $      1.58 \pm 0.02  $  &  $              1.48 \pm 0.02 $   \\ 
$\Gamma$                                                               &  $                  2.66 \pm 0.03 $  &  $         2.74^{+ 0.04}_{- 0.05} $  &  $         2.70^{+ 0.01}_{- 0.03} $  &  $                  2.66 \pm 0.03 $  &  $         2.76^{+ 0.04}_{- 0.02} $  &  $                2.76 \pm 0.03 $   \\ 
$\fsc$\tablenotemark{a}                                                &  $                0.183 \pm 0.007 $  &  $      0.21 \pm 0.01 $  &  $      0.143^{+ 0.003}_{- 0.006} $  &  $      0.180^{+ 0.009}_{- 0.006} $  &  $                0.210 \pm 0.007 $  &  $    0.168^{+ 0.011}_{- 0.007} $   \\ 
$a_{*, {\rm CF}}$\tablenotemark{b}                                     &  $      0.861^{+ 0.006}_{- 0.004} $  &  $      0.811^{+ 0.008}_{- 0.006} $  &  $                0.810 \pm 0.005 $  &  $                0.857 \pm 0.005 $  &  $                0.805 \pm 0.006 $  &  $              0.791 \pm 0.007 $   \\ 
$\Mdot$\tablenotemark{b}$ (10^{18}{\rm g~s}^{-1})$                     &  $                1.79 \pm 0.02 $  &  $         1.77^{+ 0.03}_{- 0.03} $  &  $                1.65 \pm 0.02 $  &  $                  1.81 \pm 0.03 $  &  $         1.80^{+ 0.03}_{- 0.02} $  &  $                1.69 \pm 0.02 $   \\ 
$q_1$\tablenotemark{c}                                                 &                                    3 &                                    3 &                                    3 &  $            4.4^{+ 0.2}_{- 0.4} $  &  $            4.4^{+ 0.3}_{- 0.4} $  &  $          4.9^{+ 0.2}_{- 0.3} $   \\ 
$R_{\rm br} (R_{\rm ISCO})$\tablenotemark{c}                           &                            \nodata   &                            \nodata   &                            \nodata   &  $                    3.0 \pm 0.5 $  &  $                    4.3 \pm 0.5 $  &  $                  3.7 \pm 0.4 $   \\ 
$a_{*, {\rm ref}}$                                                     &  $      0.97^{+ 0.01}_{- 0.05} $  &  $         0.82^{+ 0.06}_{- 0.27} $  &  $      0.992^{+ 0.003}_{- 0.060} $  &  $         0.84^{+ 0.03}_{- 0.07} $  &  $         0.41^{+ 0.10}_{- 0.14} $  &  $       0.94^{+ 0.01}_{- 0.08} $   \\ 
$H_{\rm den}$ (10$^{19}$~cm$^{-3}$)                                    &  $            6.4^{+ 0.4}_{- 1.2} $  &                            \nodata   &                            \nodata   &  $                    5.3 \pm 0.3 $  &                            \nodata   &                          \nodata    \\ 
$kT_{\rm ref}$ (keV)                                                   &  $      0.737^{+ 0.004}_{- 0.007} $  &                            \nodata   &                            \nodata   &  $                0.741 \pm 0.007 $  &                            \nodata   &                          \nodata    \\ 
 Illum/BB                                                              &  $   0.034^{+ 0.001}_{- 0.002} $  &                            \nodata   &                            \nodata   &  $              0.032 \pm 0.002 $  &                            \nodata   &                          \nodata    \\ 
$N_{\rm ref} (10^{-7}) / (10^{-3})$\tablenotemark{d}                   &  $                    3.3 \pm 0.3 $  &  $                  1.1 \pm 0.2 $  &                            \nodata   &  $            3.4^{+ 0.2}_{- 0.3} $  &  $            1.9^{+ 0.8}_{- 0.5} $  &                          \nodata    \\ 
$\xi$\tablenotemark{e} (erg\ cm\ s$^{-1}$)                             &                            \nodata   &  $        9500^{+ 200}_{- 1500} $  &  $       7400^{+ 1100}_{- 600} $  &                            \nodata   &  $                5800 \pm 1000 $  &  $              4100 \pm 1500 $   \\ 
$T_{\rm ref} $ ($10^6$~K)                                              &                            \nodata   &                            \nodata   &  $                    8.9 \pm 0.2 $  &                            \nodata   &                            \nodata   &  $         14.0^{+ 0.3}_{- 1.1} $   \\ 
$E_{\rm Fe} $\tablenotemark{f}(keV)                                    &                            \nodata   &                            \nodata   &  $      6.65^{+ 0.03} $  &                            \nodata   &                            \nodata   &  $    6.66^{+ 0.10}_{- 0.01} $   \\ 
$F_{\rm Fe} $\tablenotemark{f}$(10^{-4}{\rm ph~s}^{-1}\cm^{-2})$       &                            \nodata   &                            \nodata   &  $         1.0^{+ 0.1}_{- 0.4} $  &                            \nodata   &                            \nodata   &  $          1.3^{+ 0.4}_{- 0.3} $   \\ 
$\xi_{D} $\tablenotemark{e,g}  (erg\ cm\ s$^{-1}$)                     &  $         5000_{- 1500.} $  &  $                  600 \pm 200 $  &  $         900\pm 200 $  &  $           5000_{-1000}  $  &  $          540^{+ 50}_{- 150.} $  &  $      1200^{+ 200}_{- 100} $   \\ 
$N_{\rm ref, D}$\tablenotemark{g}$ (10^{-7})$                          &  $         0.50^{+ 0.14}_{- 0.06} $  &  $                      8 \pm 4 $  &  $            3.3^{+ 1.9}_{- 0.6} $  &  $                  0.5 \pm 0.1 $  &  $           12 \pm 1 $  &  $          4^{+ 2}_{- 1} $   \\ 
$N_{\rm XIS 3} / N_{\rm XIS 0}$\tablenotemark{h}                       &  $              0.960 \pm 0.002 $  &  $              0.960 \pm 0.002 $  &  $              0.959 \pm 0.002 $  &  $   0.961 \pm 0.002 $  &  $   0.960 \pm 0.002 $  &  $            0.960 \pm 0.002 $   \\ 
$N_{\rm XIS 1} / N_{\rm XIS 0}$\tablenotemark{i}                       &  $   0.915^{+ 0.002}_{- 0.001} $  &  $                0.918 \pm 0.004 $  &  $              0.917 \pm 0.001 $  &  $   0.918^{+ 0.002}_{- 0.001} $  &  $      0.916^{+ 0.005}_{- 0.003} $  &  $  0.918^{+ 0.001}_{- 0.002} $   \\
$\Delta\Gamma(XIS 1 - XIS 0) $\tablenotemark{i}                        &  $             -0.018 \pm 0.001 $  &  $               -0.015 \pm 0.003 $  &  $        -0.014 \pm 0.001 $  &  $             -0.015 \pm 0.002 $  &  $               -0.017 \pm 0.003 $  &  $           -0.016 \pm 0.001 $   \\ 
$N_{\rm PIN~CXB}$\tablenotemark{j}$(10^{-4}{\rm ph~s}^{-1}\cm^{-2})$   &  $            9.4^{+ 0.2}_{- 0.4} $  &  $                    9.1 \pm 0.6 $  &  $            9.5^{+ 0.3}_{- 0.6} $  &  $                   9.4 \pm 0.2 $  &  $          8.1^{+ 0.5}_{- 0.1} $  &  $          9.8^{+ 0.2}_{- 0.6} $   \\ 
PIN$_{\rm BG, Inst}$ Norm\tablenotemark{k}                             &  $                1.03 \pm 0.01 $  &  $         1.06_{- 0.02} $      &  $      1.041^{+ 0.004}_{- 0.006} $  &  $         1.041^{+ 0.004}_{- 0.009} $  &  $                1.05 \pm 0.01 $  &  $    1.039^{+ 0.006}_{- 0.009} $   \\ 
$N_{\rm PIN}  / N_{\rm XIS 0} $\tablenotemark{h}                       &                                 1.16 &                                 1.16 &                                 1.16 &                                 1.16 &                                 1.16 &                               1.16  \\ 
$N_{\rm RXTE}  / N_{\rm XIS 0} $\tablenotemark{h}                     &  $                0.999 \pm 0.011 $  &  $      0.989^{+ 0.006}_{- 0.010} $  &  $      1.049^{+ 0.005}_{- 0.004} $  &  $                1.001 \pm 0.011 $  &  $      0.983^{+ 0.006}_{- 0.008} $  &  $              1.060 \pm 0.005 $   \\ 
           \hline                                                       
           $\chi^2/\nu$                                                &                     898.6/1121       &                 887.2/1123           &                984.6/1131            &                    885.5/1119        &                     862.1/1121       &          944.4/1129   \vspace{-0.5mm}  \\      
\enddata
\vspace{-0.6cm} \tablecomments{ MCMC fit results.  Uncertainties are
  the minimum-width 68 per cent confidence intervals about the
  posterior mode from the MCMC runs (computed in the log for scale
  parameters), and $\chi^2/\nu$ values are listed for the best fit
  obtained.  Numbers with no uncertainty have been set to fixed
  values. We note that there are unshown fit parameters for each of
  the \rxte\ spectral fits (Models~1(iii) through 2(iv)).  In each of
  those cases, the eleven \rxte\ spectra are each fitted for an
  independent value of $\fsc$.  Additionally, for Models~1(iii),
  1p(iii), and 1(iv) and 1p(iv), each \rxte\ spectrum is fitted for
  its own value of $N_{\rm ref}$.  These values are distributed
  approximately as $\fsc=0.19\pm0.06$ and $N_{\rm
    ref}=(3\pm1.5)\times10^{-3}$ for Model~1(iii) and (iv);
  $\fsc=0.17\pm0.05$ and $N_{\rm ref}=(2\pm2)\times10^{-7}$ for
  Model~1p(iii) and (iv); $\fsc=0.15\pm0.06$ for Model~2(iii) and
  (iv).\vspace{-1mm} }

\tablenotetext{a}{The fraction of disc photons scattered into the Compton power law.\vspace{-1mm}}
\tablenotetext{b}{Continuum-fitting parameters: the spin and mass accretion rate.  Mass, inclination, and distance have been frozen at their nominal values from \citet{Gou_2009}; See \citet{Orosz_2009}. \vspace{-1mm}}
\tablenotetext{c}{Relativistic smearing indexes $q_1$ and $q_2$ describe, respectively, the inner and outer power-law illumination pattern of the disc.  The power law is broken between the two regimes at $R_{\rm br}$; $q_2$ is set to 3 in all instances.\vspace{-4mm}}
\tablenotetext{d}{Normalisation for \reflionx, in units of 10$^{-7}$, and for \refbhbm, in units of 10$^{-3}$.\vspace{-4mm}}
\tablenotetext{e}{The ionisation parameter of the reflecting material.\vspace{-1mm}}
\tablenotetext{f}{Fe-line parameters $E_{\rm Fe}$ and $F_{\rm Fe}$ give the line energy and line flux, respectively.}
\tablenotetext{g}{The subscript `D' refers to reflection emission distant from the BH, and hence not relativistically smeared (see Section~\ref{subsec:disc:distant}).\vspace{0mm}}
\tablenotetext{h}{Cross-normalisations for the detectors are referenced to XIS-0.\vspace{0mm}}
\tablenotetext{i}{The normalisation and difference in power-law index of XIS-1 are fitted with respect to XIS-0.  The resulting XIS-1 to XIS-0 cross-calibration differs in flux by $\approx5$ per cent over 1--10 keV for a Crab-like source.\vspace{-3mm}}
\tablenotetext{j}{The normalisation of the PIN's CXB spectral model. \vspace{-4mm}}
\tablenotetext{k}{The relative normalisation of the PIN's instrumental background spectrum. }

\label{tab:results}
\end{deluxetable}

\end{document}